\documentclass[aps,prl,twocolumn,groupedaddress,nofootinbib,showpacs]{revtex4}

\usepackage{amsfonts,amsmath,amssymb,bm}
\usepackage{graphicx}
\usepackage{dcolumn}

\begin{document}

\title{Chaotic and Regular Motion in Dissipative Gravitational Billiards}

\author{A. Z. G\'orski and T. Srokowski}

\affiliation{%
Institute of Nuclear Physics, PL -- 31-342 Krak\'ow, Poland }

\date{\today}

\begin{abstract}
We consider the motion of a particle subjected to the constant gravitational 
field and scattered inelasticaly by hard boundaries which possess the shape of 
parabola, wedge, and hyperbola. The billiard itself performs
oscillations. The linear dependence of the restitution coefficient on the
particle velocity is assumed. We demonstrate that this dynamical system 
can be either regular or chaotic, which depends on the billiard shape 
and the oscillation frequency. The trajectory calculations are compared with 
the experimental data; a good agreement has been achieved. Moreover, the
properties of the system has been studied by means of the Lyapunov exponents
and the Kaplan-Yorke dimension. Chaotic and nonuniform patterns visible in the
experimental data are interpreted as a result of large embedding dimension. 
\end{abstract}

\pacs{05.45.-a, 05.45.Pq, 05.45.Df}

\maketitle

From the mathematical point of view, billiards constitute an interesting class
of dynamical systems because they exhibit -- despite their simplicity -- a
variety
of nonlinear phenomena, including both regular tori and completely chaotic,
dense trajectories. Some of them are quite realistic and have a
direct physical importance. An example of such a system is the gravitational
billiard in which a point-mass particle bounces within a container of a given
shape and its motion between the bounces is not free but ballistic. Obviously,
the dynamics depends on the billiard shape. For the
parabolic boundary the system is integrable and all orbits are regular and
stable. Historically, the first study on the gravitational billiards has been
performed for the two-dimensional wedge, defined as two intersecting straight
lines \cite{leh,mil}. The motion in the wedge is fully chaotic if its vertex
angle is larger than $\pi/2$ \cite{woj}, otherwise there is a coexistence of
regular and chaotic behaviour. Those findings were successfully tested
experimentally \cite{miln}. The chaotic, as well as regular, dynamics is present
also for the hyperbolic shape of the gravitational billiard which involves both
the wedge and the parabola as its asymptotic limits \cite{fer}. The chaotic
component prevails for shapes close to the wedge. 

All the above approaches assumed that the collisions between the particle and
the
billiard boundary were elastic. It is natural to require that, in order to make
the problem more realistic, one should take into account the energy loss and
allow for the energy exchange between the
particle and the wall. Such handling of the dissipation is known in nuclear
physics as the wall formula \cite{blo}, derived in the framework of the liquid
drop
model. The atomic nucleus can then be modeled as a billiard possessing the
oscillating boundary given by the Legendre polynomials of various kinds
\cite{blo1} which lead to both regular and chaotic motion \cite{blo2}. Similar
investigations for the gravitational billiards were lacking. Only recently they
have been studied -- experimentally --
under the assumption that the energy loss during the collisions is to be
compensated, on the average, by the motion of the container \cite{fel}. The
authors of Ref.\cite{fel} constructed three aluminum containers with parabolic,
wedge, and hyperbolic shape which exercised the horizontal oscillations.
Inside a ball of steel was scattered from the boundaries and a camera
registered the position of each bounce and the ball velocity. The results
clearly indicate the regular motion for the parabola and the chaotic one
for the wedge; they also suggest some sort of regularity for the hyperbola
at a small driving frequency.

In this Letter we present a theoretical study of the inelastic gravitational
billiards with a time-dependent driving. To the best of our knowledge, it is 
the first theoretical approach to these -- very realistic -- systems. The
billiard shapes and parameters of the dynamical system have been
so chosen to enable us a direct comparison with the experimental data
\cite{fel}. Then we assume the following boundaries: $f(x)=a(x-\bar x)^2+c$ 
(the parabola), $f(x)=b|x-\bar x|+c$ (the wedge), and
$f(x)=\sqrt{\alpha(1+\beta(x-\bar x)^2)}-\delta$ (the hyperbola), where $a=0.26$
cm$^{-1}$, $b=1.85$, $c=0.63$ cm, $\alpha=40.3$ cm$^2$, $\beta=0.08$ cm$^{-2}$,
and $\delta=4.45$ cm. The containers oscillate horizontally: 
$\bar x(t)=-A\sin \omega t$, 
where $A$ is the amplitude and $\omega=2\pi f$ is the oscillation
frequency. Inside the container the particle is subjected to the constant
gravitational acceleration $g$. Collisions with the boundaries result in the
energy loss, quantified by the restitution coefficient $r\in [0,1]$ which is
defined as a ratio of the absolute values of the velocity after and before the
collision. The case $r=1$ corresponds to the elastic collision. It is
difficult to decide {\it a priori} which value for $r$ should be assumed. An
experiment with steel particles bounced on a steel block gives $r=0.93$
\cite{kud}. However, taking into account effects connected with the sharing of 
energy between rotation and translation during the collision reduces this
coefficient substantially and the effective $r$ appears of about 0.7. Moreover,
$r$ can depend on the velocity and on the scattering angle \cite{kud}. The
authors of Ref.\cite{fel} suggest $r=0.9$. In the following, we will try to
draw some conclusions about the restitution coefficient from the comparison
of our results with the experimental data of Ref.\cite{fel}.

Let us assume that the particle hits the boundary at the time $t_0$ with the
velocity ${\bf v}^C_0$, determined in respect to the frame connected
with the billiard, and the collision point is $(x^C_0,y^C_0=f(x^C_0))$. 
The transformation of particle velocities at this point, 
${\bf v}^C_0\to {\bf v}^C_1$, has the following form
\begin{equation}
\label{trvel}
{\bf v}^C_1=r({\bf v}^C_0-2{\bf u}({\bf v}^C_0\cdot{\bf u})),
\end{equation}
where the components of the versor normal to the boundary, ${\bf u}$, depend on 
$x^C_0$ and are given by: $u_x=-f'(x^C_0)/h(x^C_0)$ and $u_y=1/h(x^C_0)$ with
$h(x^C_0)=\sqrt{1+f'^2(x^C_0)}$. The particle, after being reflected from the
boundary, moves along the ballistic trajectory:
 \begin{equation} 
 \label{bal}
  \begin{split}
  & x^C(t)=x^C_0+v^C_{x,1}(t-t_0)+A\sin\omega t-A\sin\omega t_0 \\
  & y^C(t)=y^C_0+v^C_{y,1}(t-t_0)-g(t-t_0)^2/2,
  \end{split}
\end{equation}
for $t>t_0$, up to the next section of this curve with the boundary:
$(x^C(t_1),y^C(t_1))=(x^C_1,y^C_1)$. 
The subsequent applying of Eqs.(\ref{trvel}) and (\ref{bal}) produces a set of
collision events which take place at times $t_n$. 
The time evolution can be characterized by the vector 
${\bf X}_n=(x^C_n,y^C_n,v^C_{x,n},v^C_{y,n})$ in the four-dimensional 
phase space, where the velocities are taken just before the consecutive 
bounces. Therefore, we can restrict the dynamics to the billiard
boundary and describe it in terms of the following time-dependent mapping:
\begin{equation}
\label{map}
{\bf X}_{n+1}={\cal P}_n({\bf X}_n).
\end{equation}
The above expressions have been formulated in the billiard coordinates 
because then the simple velocity transformation rule (\ref{trvel}) holds.
The transformation to the laboratory frame is straightforward:
$x=x^C-A\sin\omega t$ and $v_x=v^C_x-A\omega\cos\omega t$; $y$-components
remain the same. The equation for the boundary $f(x)$ reduces the 4-dimensional
phase space, in which ${\cal P}_n$ is defined, to the 3-dimensional manifold.

For various shapes of the billiard and various driving forces, the mapping
${\cal P}_n$ can represent either a regular cycle or a strange attractor. The
static, elastic billiard with the parabolic shape is always regular and 
possesses two stable orbits: the horizontal orbit, 
connected with the symmetric bouncing between the
left and the right part of the boundary at the same height, and the vertical 
one which involves the top of the parabola \cite{kor}. In our case 
the limit cycle corresponds to the fixed point of the mapping ${\cal P}_n^2$ and
can be obtained analytically by solving the equation 
${\bf X}^*={\cal P}^2({\bf X}^*)$. The detailed equations are complicated
and
will not be presented here. As a result, we yield the stable horizontal orbit 
which moves to and fro together with the container. The vertical orbit of the
elastic billiard shrinks now to a single point. The time interval between
consecutive bounces of the horizontal orbit (2-point attractor on the boundary) 
$\tau^*=\pi/\omega$, moreover $v_y=\pi g/2\omega$; 
these quantities depend neither on the restitution coefficient $r$ 
nor on the billiard shape. For the parabolic shape the agreement of $\tau^*$
with the data \cite{fel} is very good. For $r=0.9$ we get the
height of the orbit $y=158$ cm which exceeds by far the experimental value
$y\approx 7.5$ cm. The latter value can be obtained if we assume $r=0.43$. 
Therefore, the restitution coefficient seems to be well established by the
experiment at the value $r=0.43$ and we can try to
apply it in numerical calculations for the other shapes. However, 
all trajectories calculated in this way, both for the wedge and for the
hyperbola, do collapse to the bottom of the billiard instantly. 
Apparently, the assumption that the restitution coefficient 
can be approximated by a constant cannot be maintained in the present problem. 

In the following we assume that the damping factor depends linearly on the
particle velocity and then the restitution coefficient is of the form:
\begin{equation}
\label{reco}
r=1-|{\bf v}^C|/v_{cr}
\end{equation}
if $r>0.01$; otherwise (for very large velocities) $r=0.01$. The parameter
$v_{cr}$ can be determined by comparison with the data. The parabola case is
especially useful for that purpose because it must be characterized by a simple
cycle. The data \cite{fel} seem to confirm the presence of a cycle though it
appears in a
strongly diffused form which may result both from the low resolution in the
experiment (e.g. due to the surface roughness) and from the rotational degree of
freedom, not taken into account in the calculations. We get the position of the
2-point attractor in agreement with the data for $v_{cr}=390$ cm/s. We apply
this value in the calculations for the other shapes. 

\begin{figure}
  \includegraphics[width=8.2cm,angle=0]{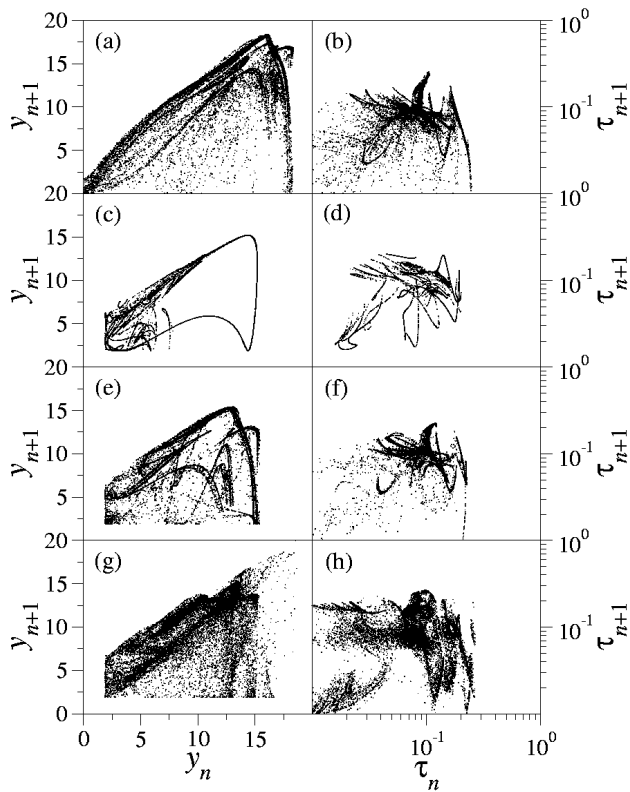} 
  \caption{The coordinate $y$ of a collision point vs. $y$ for the previous
collision (left column) and the time of flight between two subsequent collisions
(right column). The following cases are presented: the wedge with $f=6.6$ Hz
((a) and (b)) and the hyperbola with: $f=4.5$ Hz ((c) and (d)), $f=5.8$ Hz
((e) and (f)), and $f=8$ Hz ((g) and (h)). The units are: [cm] for $y$ and [s]
for the time.}
\label{fig:andrzej1}
\end{figure}

Fig.1 presents some properties of the map ${\cal P}_n$ -- the height $y$
and the time interval between consecutive bounces $\tau_n=t_n-t_{n-1}$ -- for
the wedge and the
hyperbola which is driven by three different frequencies. Each figure
represents a single trajectory, evolved up to $t=2\cdot10^3$ s which corresponds
to about $2\cdot10^4$ collisions with the boundary. The plot of $y$
for the wedge is strongly nonuniform and indicates 
a high degree of chaoticity.
The most of the points is concentrated just above the line $y_{n+1}=y_n$ and
the fractal structure is hardly visible. The
figure can be directly compared with the data (see Fig.3 in Ref.\cite{fel}); 
the similarity is striking though the calculated quantities are
extended to slightly larger values than the data. The nonuniformity is clearly
visible also in the plot of the time intervals and the region close to the
point $(0.09,0.09)$ is distinguished. The experimental data exhibit the same
pattern. The hyperbola involves the wedge as well as the parabola as
its limiting shapes and one can expect both shapes influence the results of the
dynamical calculations for the hyperbola. 
For the low driving frequency $f=4.5$ Hz (Fig.1c, d) the
particle abides close to the bottom of the billiard which can be well
approximated by the parabola. That results in apparently regular pattern.
However, a stochastic ingredient, similar as in Fig.1a which corresponds
to the wedge, is also visible in Fig.1c. In the experimental data the chaotic
component connected with the wedge
seems to be absent completely. Instead, long-time tails in the plot of time
intervals, possessing the intermittent structure, are observed, as well as very
small values of the height $y$. Then the experimental results suggest a more
regular motion than the calculations predict. For the larger frequency 
$f=5.8$ Hz (Fig.1e, f) the chaotic behaviour is overwhelming. The bands
typical for the strange attractors are visible; it is not the case for the
experimental data but such subtle structures may be smeared due to the low
resolution. The bands vanish completely if we make the
frequency still larger. Fig.1 g, h shows that it happens for $f=8$ Hz (the case
not studied experimentally) and the picture is similar to that for the
wedge. However, it is not generally true that the degree
 of chaoticity rises with the frequency: for $f>8$ Hz the motion
becomes regular again. 

\begin{figure}
  \includegraphics[width=8.2cm,angle=0]{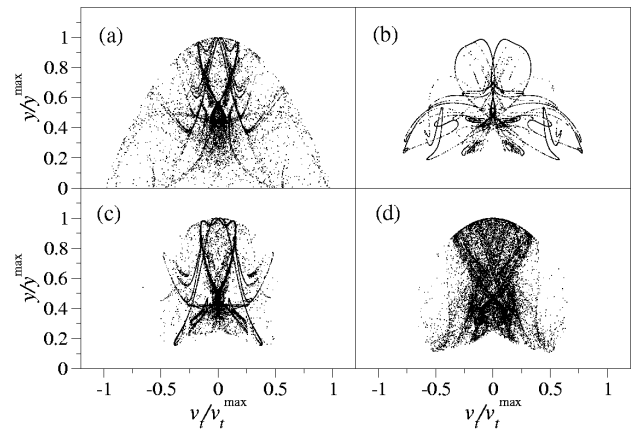}%
  \caption{The normalized coordinate $y$ of a collision point vs. the
normalized tangential velocity $v_t$ at the same point for the following cases:
the wedge with $f=6.6$ Hz (a) and the hyperbola with $f=4.5$ Hz (b), with
$f=5.8$ Hz (c), and with $f=8$ Hz (d).}%
\label{fig:andrzej2}%
\end{figure}

Some additional information about the phase space structure can be obtained by
plotting the position $y$ versus the tangential velocity after the
collision, $v_t$, at collision points, normalized by $y^{max}$ and $v_t^{max}$,
respectively. The quantities $y^{max}$ and $v_t^{max}$ mean the largest possible
values of $y$ and $v_t$ at each collision, obtained under the assumption that
the particle is either at rest or bounces almost horizontally at the bottom of
the billiard. Therefore, the region around $v_t=0$ corresponds to the head-on
collisions which may result in instability of the motion and the onset of the
chaotic behaviour. For the parabolic shape we have obtained 2-point limiting
cycle, corresponding to very small values of the tangential velocity:
$(v_t/v_t^{max},y/y^{max})=(\pm0.044,0.39)$. The experimental result presents
itself as a narrow band, due to the noise in the system. Results of the
calculations for the other shapes are presented in Fig.2. The case of the
hyperbola for the largest frequency $f=8$ Hz shows the greatest disorder
(Fig.2d) whereas the trajectory for the wedge (Fig.2a) is able to fill the
entire region which is allowed by the energy partition rule \cite{fel}. 
The figure for the hyperbola at the intermediate frequency $f=5.8$ Hz 
(Fig.2c) indicates, in turn, a pronounced fractal structure. 
On the other hand, the pattern for $f=4.5$ Hz is predominantly regular
(Fig.2b). However, a vertical strip at small $|v_t|$ bears apparent
signs of chaos. Indeed, a magnification of this region reveals the fractal
structure. Also the experiment \cite{fel}
distinguishes the region of small tangential velocities for this case but the
lack of any fine structure in the data prevents detailed comparisons.

\begin{table}
  \caption{\label{tab:table1} The Liapunov exponents for the mapping 
  ${\cal P}_n$ and the Kaplan-Yorke
dimension for the parabola (p), the wedge (w), and the hyperbola with
frequencies: f=4.5 Hz (h1), f=5.8 Hz (h2), and f=8 Hz (h3).}
  \begin{ruledtabular}
    \begin{tabular}{cddddd}%
      \ & \lambda_1 & \lambda_2 & \lambda_3 & \lambda_4 & D_{KY}\\
      \hline%
        p & -0.408 & -0.579 & -0.579 & -2.17 & 0.00 \\
     \ \ \ w  & 0.447  & -0.288 & -1.17 & -1.98  & 2.14 \\
     \ \ \ h1 & 0.208  & -0.579 & -1.05 & -1.42 & 1.36 \\
     \ \ \ h2 & 0.367 & -0.394 & -1.21 & -1.78 & 1.93 \\
     \ \ \ h3 & 0.611 & -0.0637 & -1.05 & -2.08 & 2.52 \\
    \end{tabular}
  \end{ruledtabular}
\end{table}
A picture which has emerged so far is not complete yet because we do not have
any quantitative knowledge -- like the degree of instability and the
attractor dimensionality -- about the chaotic motion visible in the figures.
Then we calculate the Liapunov exponents for the mapping ${\cal P}_n$ which
characterize the evolution of
dynamical systems in the tangent space. Since the billiard is not smooth and
the direct linearization of the equations of motion is not possible, we
resort to an approximate method \cite{ben,ben1} 
by utilizing the fact that the distance between
two close trajectories is governed by the linearized equations. Then we perform
the time evolution of two trajectories, whose initial conditions differ by 
$\delta r$, by the time interval $\delta t$ which must be small enough to keep
the trajectories close. In the next step we renormalize the relative distance
to $\delta r$ and continue the procedure for a long time. In order to get the
entire spectrum of four Liapunov exponents $\lambda_i$ we
need to evolve five trajectories and utilize the Gramm-Schmidt
orthogonalization scheme at each renormalization step to sort manifolds 
connected with subsequent unstable and stable directions \cite{shi}. Finally,
 we have to multiply the exponents by the mean time between successive
bounces, different for each case. The
results are summarized in Table I. All presented cases are characterized by one
positive exponent, except the parabola for which all exponents are negative. For
the hyperbola, the instability rises with the frequency and even the
trajectory for the smallest frequency is chaotic despite the apparent regularity
visible in the figures.

Having all Liapunov exponents calculated, we can determine the Kaplan-Yorke
dimension which in many cases may be identified with the Hausdorff dimension
(the Kaplan-Yorke conjecture \cite{pei}). The definition is the following
\begin{equation}
\label{ky}
D_{KY}=j+\frac{\sum_i^j\lambda_i}{-\lambda_{j+1}},
\end{equation}
where $j$ is the largest integer such that
$\lambda_1+\lambda_2+\dots+\lambda_j>0$. The results, presented in Table I,
indicate that the embedding dimension of the attractor for the cases $w$ and
$h3$ equals 3, i.e. it is equal to the entire available manifold. This
conclusion explains the nature of chaotic, but also nonuniform and diffused,
pattern observed in the figures for those cases, 
in respect both to the theoretical predictions and to
the experimental data: for so high dimensionality of the attractor its
structure cannot be clearly visible in the plane. For the wedge that diffused,
nonfractal pattern is restricted to the area just above the line $y_{n+1}=y_n$
but in the case $h3$ it extends to the almost whole picture.
On the other hand, the fractal structure is apparent for the case $h2$. 

We have demonstrated that the inelastic gravitational billiard with the
time-dependent 
driving, possessing the shape of the wedge and the hyperbola, constitutes the
strange attractor, whereas the parabolic shape is characterized by the regular
motion. We have proposed a simple dependence of the restitution coefficient
on the velocity. Such dependence appears to be essential to get results
consistent with the experimental data. Some complicated patterns, revealed by the experiment, 
can be explained by the attractor dimensionality and its dependence on the
billiard shape and parameters. Generally, the model predictions agree 
quite well with the data. However, the values of the
height $y$ are slightly too
large, which results in the more irregular motion than the experiment shows, 
for the hiperbolic shape with small driving frequency. This discrepancy may be
a token of too weak damping and a suggestion that the simple formula 
(\ref{reco}) should be refined at small velocities. For that purpose some
experimental effort is
necessary: the resolution of the data should be improved and, first of all, the
restitution coefficient, which is the essential quantity in the model, should
be determined in the wide range of the ball velocity.

\end{document}